# EasyMergeR: an interactive Shiny application to manipulate multiple XLSX files of multiple sheets

**Ziyu Zhu[1], Ximing Xu[1,2,*]**

[1] Big Data Center for Children's Medical Care, Children's Hospital of Chongqing Medical University, Chongqing 400014, China

[2] National Clinical Research Center for Child Health and Disorders, Ministry of Education Key Laboratory of Child Development and Disorders, Chongqing 400014, China

[*] Correspondence: ximing@hospital.cqmu.edu.cn

## Abstract

**Summary:**

The integration of sequencing data with clinical information is a widely accepted strategy in bioinformatics and health informatics. Despite advanced databases and sophisticated tools for processing omics data, challenges remain in handling the raw clinical data (typically in XLSX format with multiple sheets inside), either exported from health information system (HIS) or manually collected by investigators. This is particularly difficult for time-constrained medical staff with little or no programming background, and it is typically the first bottleneck in many clinical-oriented studies. To fill this gap, we developed EasyMergeR, a simple, user-friendly, code-free R Shiny application that allows interactive manipulation of multiple XLSX files with multiple sheets and provides basic data manipulation capabilities based on the tidyverse and other handy R packages.

**Availability and implementation:**

EasyMergeR is implemented in R and Shiny and is freely available on GitHub (https://github.com/ai4kids/EasyMergeR). This package can be installed via the devtools R package[1]. EasyMergeR is also accessible through the shinyapps.io service (https://dodc.shinyapps.io/easymerger/).

# 1. Introduction

We have witnessed an unprecedented explosion of health information in the past decades due to rapid development of digital technologies and even more robust infrastructures[2]. The widely-adopted hospital information system (HIS) has integrated a wealth of health information from disparate repositories, such as electronic medical record (EMR) system, pharmacy information system (PIS), laboratory information system (LIS), electronic health record (EHR) system and so forth. Consequently, HIS and systems alike have made it possible and feasible to conduct studies using various health information data in accordance with relevant regulations[3]. However, the potential of this gold mine of information is far from being fully realized partially due to the lack of engagement from end users like doctors[4]. One important reason for this gap is that although doctors are proficient with their clinical expertise and life-saving skills, many are unfamiliar with coding and programming, adding to the difficulties of kicking off their research, which usually requires excruciating data preparation by either gleaning data from piles of medical records or exporting excessive raw data from their hospitals' databases.

Generally, as highly integrated information systems, HIS and other systems support exporting copious data in XLSX or CSV (convertible to XLSX) files so that doctors can prepare data directly using the Excel software. However, HIS usually yields multiple workbooks with multiple sheets inside rather than one simple and straightforward table as is expected. Processing such data can be difficult and painful for doctors who lack programming skills. In the R community, many handy R packages have been developed to process XLSX files like openxlsx and readxl, and subsequent data manipulation is also made easy thanks to the tidyverse package ecosystem. Nevertheless, these tools require somehow familiarity or a basic understanding of R programming language. To this end, tutorials and seminars are undoubtedly helpful. However, an intuitive, interactive, and code-free software might be easier and more helpful for users like doctors considering the fact that they are already exhausted in heavy daily clinical workloads and teaching tasks.

Herein, we present EasyMergeR for manipulation of multiple XLSX files and basic data manipulation with ease based on the R Shiny framework. After uploading the XLSX files, the desired table should be readily available within a few clicks. We hope that EasyMergeR can assist in the data preparation process and help users proceed with their research.

## 2. Methods

EasyMergeR was developed in R (version 4.1.1 or newer) and Shiny [5] via the golem [6] framework for building modular and robust R packages and Shiny applications. EasyMergeR was made possible by wrapping up the functions of openxlsx [7] R package and tidyverse ecosystem R packages: dplyr[8], tidyr[9], stringr [10], and lubridate [11]. skimr [12] R package was applied for table information summary. The philosophy of EasyMergeR development consists of a two-step approach: (1) load and prepare data; (2) manipulate and export the table. Current EasyMergeR contains two modules: Module1 for file upload and merging and Module2 for basic data manipulation and table export. The R package DT[13] was applied for data table rendering. The modules and user interface of EasyMergeR are shown in **Figure.1**.

## 3. Results

A complete EasyMergeR workflow consists of the following six steps: (1) file upload; (2)

select target sheets and columns; (3) choose the merge method; (4) preview table and transform column type; (5) data manipulation; (6) export table. Briefly, the users first upload single or multiple XLSX files (unique sheet names required), and it would take several minutes or more to read large XLSX files. Next, users can freely select desired sheets and columns for merging operations. We offered three merging options: "column bind", "row bind", and "no bind". Before executing the merge operation, EasyMergeR would perform series of checks to make sure that the merge option is correct and prompt the user for confirmation whenever the merge operation might result in a large table, which usually means an inappropriate merge configuration. The merged data would be sent to the second module for further processing after clicking the "Next" button. The current version of EasyMergeR provides basic data manipulation functionality and sorts them into two categories: column operations and row operations. In case of wrong or unwanted operations, EasyMergeR supports saving and restoring the working table. Finally, users select the desired columns into a table and export it in XLSX format.

## 4. Conclusion

The EasyMergeR R shiny application provides the ability to interactively merge multiple XLSX files and perform basic data manipulation for all users, especially those without coding experience. EasyMergeR facilitates the processing of raw clinical information through a simple and interactive user interface, and hopefully encourages research involving clinical information mining.

**Funding**

This work was supported by grants from the Chongqing Innovation Program for Returned Overseas Chinese Scholars (ex2021112) and the CQMU Program for Youth Innovation in Future Medicine (No. W0124).

**Figure legends**

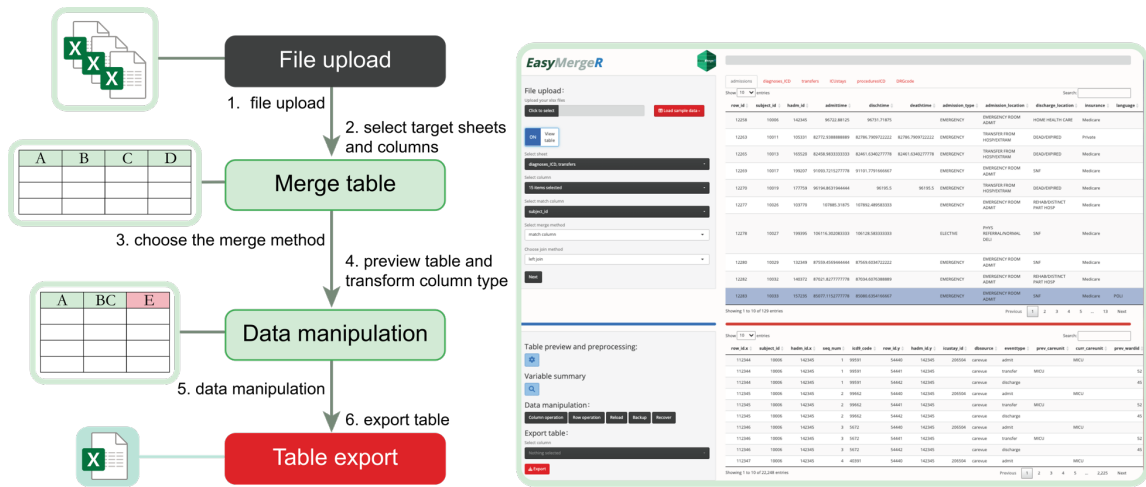

**Figure.1 Screenshot of EasyMergeR and workflow diagram.**

The workflow of EasyMergeR consists of four steps. Starting from the file upload, users can next merge the table and send it for data manipulation next. Finally, users can select all the columns as the final table to export.